\documentclass[namedreferences]{solarphysics}
\usepackage[hyperref,optionalrh,solaromanenum]{spr-sola-addons} 
\usepackage{graphicx}                    
\usepackage[usenames, dvipsnames]{color}
\usepackage{breakurl}                         

\begin{document}

\begin{article}
\begin{opening}

\title{Long-Term Tracking of Corotating Density Structures using Heliospheric Imaging}
%
 \author[addressref={aff1,aff2},corref,email={iplotnikov@irap.omp.eu}]{\inits{I.~Plotnikov}\fnm{I.~Plotnikov}} \sep 
 \author[addressref={aff1,aff2},corref,email={arouillard@irap.omp.eu}]{\inits{A.P.~Rouillard} \fnm{A.P.~Rouillard}} \sep 
 \author[addressref={aff3}]{\inits{J.A.~Davies} \fnm{J.A.~Davies}} \sep
 \author[addressref={aff4}]{\inits{V.~Bothmer} \fnm{V.~Bothmer}} \sep
 \author[addressref={aff5}]{\inits{J.P.~Eastwood} \fnm{J.P.~Eastwood}} \sep
 \author[addressref={aff9}]{\inits{P.~Gallagher} \fnm{P.~Gallagher}} \sep 
 \author[addressref={aff3}]{\inits{R.A.~Harrison} \fnm{R.A.~Harrison}} \sep
 \author[addressref={aff8}]{\inits{E.~Kilpua} \fnm{E.~Kilpua}} \sep 
 \author[addressref={aff6,aff7}]{\inits{C.~Mostl} \fnm{C.~M\"ostl}} \sep  
 \author[addressref={aff3}]{\inits{C.H.~Perry} \fnm{C.H.~Perry}} \sep  
 \author[addressref={aff10}]{\inits{L.~Rodriguez} \fnm{L.~Rodriguez}} \sep
 \author[addressref={aff1,aff2}]{\inits{B.~Lavraud} \fnm{B.~Lavraud}} \sep
  \author[addressref={aff1,aff2}]{\inits{V.~G\'enot} \fnm{V.~G\'enot}} \sep
   \author[addressref={aff1,aff2}]{\inits{R.F.~Pinto} \fnm{R.F.~Pinto}} \sep
    \author[addressref={aff1,aff2}]{\inits{E.~Sanchez-Diaz} \fnm{E.~Sanchez-Diaz}}
\address[id=aff1]{Institut de Recherche en Astrophysique et Plan\'etologie, Universit\'e de Toulouse (UPS), Toulouse, France}
\address[id=aff2]{Centre National de la Recherche Scientifique, UMR 5277, Toulouse, France}
\address[id=aff3]{RAL Space, STFC Rutherford Appleton Laboratory, Harwell Campus, Didcot, OX11 0QX, UK}
\address[id=aff4]{Georg-August-Universit\"at G\"ottingen, G\"ottingen, Germany}
\address[id=aff5]{Imperial College London, London, SW7, United Kingdom}
\address[id=aff6]{Space Research Institute, Austrian Academy of Sciences, A-8042 Graz, Austria}
\address[id=aff7]{IGAM-Kanzelh\"ohe Observatory, Institute of Physics, University of Graz, A-8010 Graz, Austria}
\address[id=aff8]{University of Helsinki, Helsinki, Finland}
\address[id=aff9]{Trinity College Dublin, Dublin, Ireland}
\address[id=aff10]{Royal Observatory of Belgium, Brussels, Belgium}
%
\runningauthor{I.~Plotnikov et al.}
\runningtitle{CDS catalogue 2007-2014}

\begin{abstract}
The systematic monitoring of the solar wind in high-cadence and high-resolution heliospheric images taken by the \textit{Solar-Terrestrial Relation Observatory} (STEREO) spacecraft permits the study of the spatial and temporal evolution of variable solar wind flows from the Sun out to 1~AU, and beyond. As part of the EU Framework 7 (FP7) Heliospheric Cataloguing, Analysis and Techniques Service (HELCATS) project, we have generated a catalogue listing the properties of 190 corotating structures well-observed in images taken by the \textit{Heliospheric Imager} (HI) instruments on-board STEREO-A (ST-A). Based on this catalogue, we present here one of very few long-term analyses of solar wind structures advected by the background solar wind. We concentrate on the subset of plasma density structures clearly identified inside corotating structures. This analysis confirms that most of the corotating density structures detected by the heliospheric imagers comprises a series of density inhomogeneities advected by the slow solar wind that eventually become entrained by stream interaction regions. We have derived the spatial-temporal evolution of each of these corotating density structures by using a well-established fitting technique. The mean radial propagation speed of the corotating structures is found to be $311 \pm 31$ km~s$^{-1}$. Such a low mean value corresponds to the terminal speed of the slow solar wind rather than the speed of stream interfaces, which is typically intermediate between the slow and fast solar wind speeds ($\sim$400 km~s$^{-1}$). Using our fitting technique, we predicted the arrival time of each corotating density structure at different probes in the inner heliosphere. We find that our derived speeds are systematically lower by $\sim 100$ km~s$^{-1}$  than those measured \textit{in situ} at the predicted impact times. Moreover, for cases when a stream interaction region is clearly detected \textit{in situ} at the estimated impact time, we find that our derived speeds are lower than the speed of the stream interface measured \textit{in situ} by an average of 55 km~s$^{-1}$ at ST-A and 84 km~s$^{-1}$ at STEREO-B (ST-B). We show that the speeds of the corotating density structures derived using our fitting technique track well the long-term variation of the radial speed of the slow solar wind during solar minimum years (2007--2008). Furthermore, we demonstrate that these features originate near the coronal neutral line that eventually becomes the heliospheric current sheet. 

\end{abstract}

%

\end{opening}

\section{Introduction} \label{s:introduction}

\indent Solar cycle variations of the solar wind have been studied via \textit{in situ} measurements made beyond 0.3~AU \citep[\textit{\textit{e.g.}} OMNI; ][]{2005JGRA..110.2104K}, coronagraphic imagery \citep{1994JGR....99.4201W} and interplanetary scintillation \citep{2010SoPh..261..149B}. Past studies have shown that the origin and evolution of interplanetary plasma vary greatly over the course of the solar cycle. At solar minimum, the solar wind is largely controlled by large-scale coronal holes generating fast solar wind whereas at solar maximum, the interplanetary plasma output and its variability are strongly modified by the increasing number of coronal mass ejections (CMEs). \\ 

\indent The large-scale structure of the solar wind measured in the ecliptic plane at solar minimum is, to a large extent, set by the recurring compression/rarefaction regions formed by the radial alignment of fast and slow solar wind. The compression regions are called stream interaction regions (SIRs) when measured once \textit{in situ} or corotating interaction regions (CIRs) when measured over consecutive solar rotations \cite[]{2006SoPh..239..337J}. It should be noted that interaction regions can also form ahead of fast CMEs that compress the interplanetary medium, occasionally generating a sheath/shock system.  

\indent On 26 October 2006 UT, NASA launched its STEREO mission \citep{2005AdSpR..36.1483K, 2008SSRv..136....5K}, which consists of two near-identical spacecraft. Each spacecraft used close flybys of the Moon to escape into near 1~AU heliocentric ecliptic orbit, with one spacecraft trailing Earth (ST-\textit{Behind} or ST-B) and the other leading Earth in its orbit (ST-\textit{Ahead} or ST-A). The \textit{Heliospheric Imager} \citep[HI; ][]{2009SoPh..254..387E} instruments on each STEREO spacecraft provide white-light imaging of the inner heliosphere, via Thomson scattering, which allows us to track continuously the solar wind outflow from $\sim 15$ $R_\odot$ from the Sun out to 1~AU, and beyond. This enables us to study the origin of the variable plasma output from the Sun. Heliospheric imaging from STEREO has shown that, contrary to the standard picture of a smooth spiral of enhanced density, SIR/CIRs have significant longitudinal variability associated with the continual release and subsequent compression of small-scale transients in the slow solar wind  \citep[\textit{\textit{e.g.}}, ][]{2010JGRA..115.4103R}.  STEREO has been monitoring the variable plasma output of the Sun systematically over much of the last decade, thereby offering just under a solar cycle of insightful observations. \\

\indent The purpose of this paper is three-fold: (1) to introduce a new catalogue of corotating density structures (CDSs) derived from ST-A heliospheric imagery (for the period covering April 2007 to August 2014), (2) to demonstrate that under quiet and moderately active solar wind conditions, heliospheric imaging off the East limb of the Sun (from the spacecraft vantage point) can be used to track systematically the 3-dimensional (3D) evolution of SIR/CIRs but that, at solar maximum, CDSs are less visible due to the increased occurrence of CMEs, (3) to study the change in the occurrence of CDSs over the solar cycle in response to the changing topology of the coronal magnetic field.

\section{Methodology for tracking corotating density structures}
\label{s:methodology}

Along with a comprehensive complement of \textit{in situ} instrumentation, each STEREO spacecraft carries a suite of imagers - the SECCHI package \citep{2008SSRv..136...67H}. SECCHI comprises an \textit{Extreme Ultraviolet Imager} (EUVI), two coronagraphs (COR-1 and COR-2), and two \textit{Heliospheric Imagers} (HI-1 and HI-2). The HI instruments are described in detail by \cite{2009SoPh..254..387E}, \cite{2009SoPh..256..219H} and \cite{2009SoPh..254..185B}. The detectors have image-pixel size of 70 arcsec (HI-1) and 4 arcmin (HI-2), observing in white-light with a pass band of 630-730 nm and 400-1000 nm, respectively. The HI images are recorded  at $2048\times2048$ pixels (images are binned onboard to $1024\times1024$ pixels). The images are taken every 40 min for HI-1 and every 2~hours for HI-2. 
 The HI-2 camera records images with the lowest time cadence of the two HI cameras but has the widest field of view  extending over 70$^o$ and centered at 53.7$^o$ elongation.\\

 To track individual features as they propagate through the fields of view of the heliospheric imagers, maps of brightness variation are often created by extracting bands of pixels along a
constant position angle (PA), corresponding to a fixed solar radial, and displaying them as a function of elongation (Y-axis) and time (X-axis). Such time--elongation maps are often referred to as J-maps \citep{1999JGR...10424739S, 2008ApJ...675..853S, 2008ApJ...674L.109S, 2009GeoRL..36.2102D}. For a spacecraft located in the ecliptic plane, selection of a slowly time-varying PA is required to track plasma structures propagating in the ecliptic. The resulting image bin angular sizes of HI-2 are 4.5 arcmins but the bins used to produce the J-maps overlap so that a single J-map bin covers an  elongation range of 9 arcmins.\\

Most studies published to date construct J-maps from running-difference images thereby mainly revealing the motion of density structures. In J-maps constructed from ST-A HI images, characteristic patterns of converging tracks appear during nearly every solar rotation; these, as demonstrated later in this paper, are most clearly visible during solar minimum years. As previously demonstrated by \cite{2008GeoRL..3510110R} and \cite{2008ApJ...674L.109S}, each track in any such pattern corresponds to the Thomson-scattered white-light signature of a strong density inhomogeneity (a so-called `density blob') moving radially outward from the Sun. Beyond about $30~R_\odot$ (situated roughly in the overlap region between the HI-1 and HI-2 fields of view), these inhomogeneities become entrained ahead of corotating high-speed streams \citep{2009ApJ...702..862T, 2010JGRA..115.4103R}. Because these density structures are emitted by a spatially limited source region on the Sun, they rapidly form a spiral of density inhomogeneities in the interplanetary medium. This spiral is analogous to the Parker spiral formed by the interplanetary magnetic field, as both trace approximately the locus of plasma emitted by a single co-rotating source region on the Sun. The mechanism for the release/ formation of these density inhomogeneities is not yet fully understood. Imagery shows that it is a quasi-periodic process occurring on a timescale of some 8-12 hours \citep{2008GeoRL..3510110R}. It has, however, been suggested that they are formed near the Sun inside helmet streamers through magnetic reconnection \citep{2000JGR...10525133W, 2005ApJ...625..463L, 2010JGRA..115.4103R}. A handful of these density inhomogeneities have been continuously tracked outward to spacecraft making \textit{in situ} measurements near 1~AU where they have been identified as being associated with twisted magnetic fields (perhaps magnetic flux ropes), distinct from the `quiet' interplanetary magnetic field \citep{2010JGRA..115.4104R, 2009SoPh..256..327K}. \textit{in situ} measurements also show that these twisted magnetic fields become entrained and compressed by high-speed streams before they reach 1~AU \citep{2010JGRA..115.4104R}. For the reasons outlined in the next paragraphs, we make no \textit{a priori} assumption about the nature of these density inhomogeneities. We focus on density structures that are part of a pattern of converging tracks in ecliptic J-maps generated from ST-A heliospheric imaging observations. We refer to such patterns as `corotating density structures' (CDS). We will show that, in most cases, the CDSs follow closely the evolution of the heliospheric plasma sheet. Identification of CDSs from ST-B is more difficult due to instrumental reasons that will be discussed later. \\

  \begin{figure}  
   \centerline{\hspace*{0.015\textwidth}
               \includegraphics[width=\textwidth,clip=]{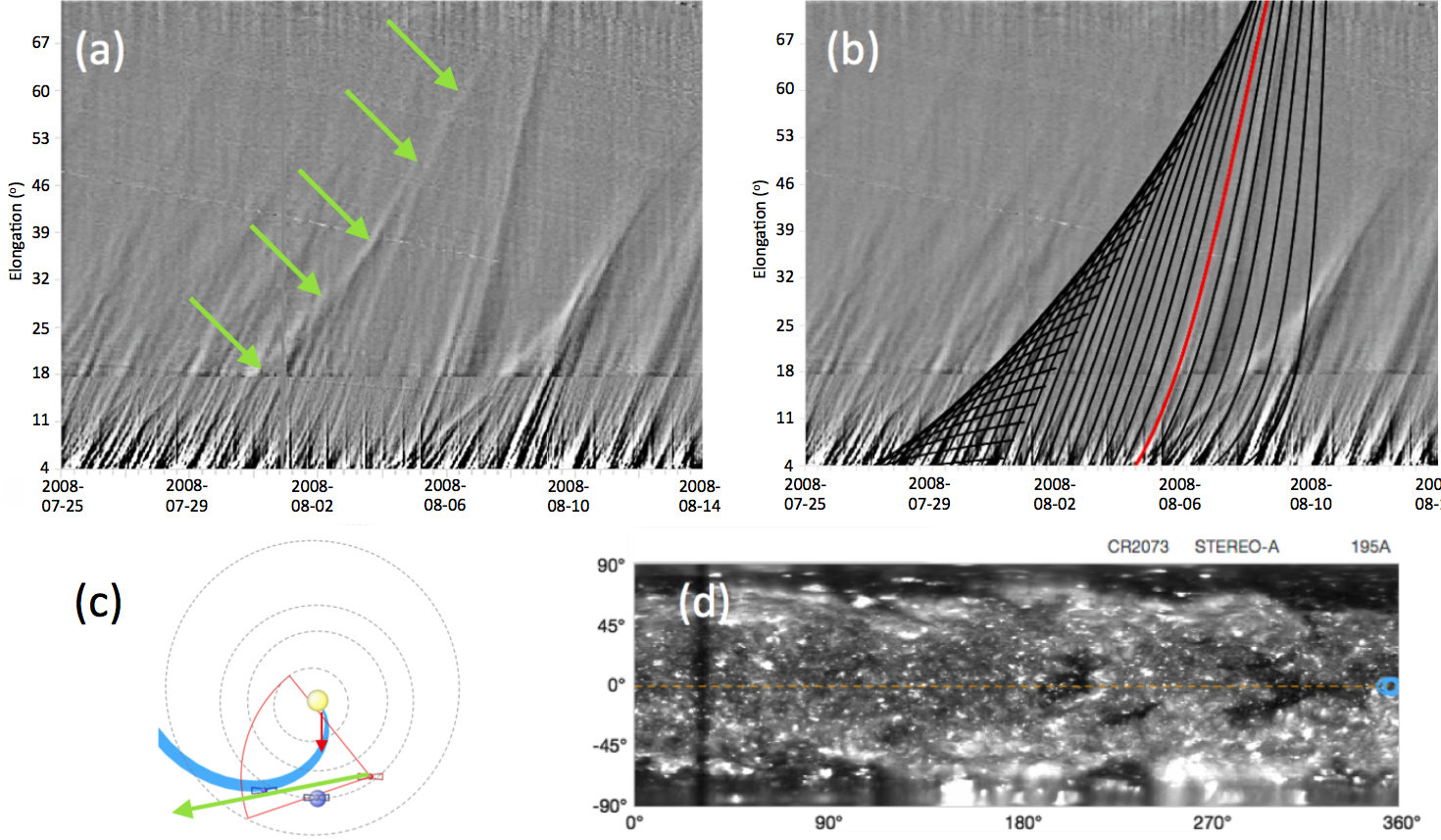}
              }
\caption{(a) Example of a J-map extending from 25 July  to 14 August 2008. (b) The same J-map but overlaid with time--elongation profiles corresponding to a series of individual blobs that comprise a CDS. The red curve is the actual fit to one well observed CDS blob (emission time 2008-08-03T21:36:26 UT, $V_b =358 \pm 10$ km~s$^{-1}$ and $\phi = $34$^{o}$ $\pm$ 3$^{o}$). The black curves, reconstructed using this speed, simulate the elongation variations of a series of such blobs emitted at 8-hour intervals from the same region on the Sun. (c) The orbital configuration at the emission date (see detailed comments in text). (d) The coronal map for the Carrington rotation 2073 at a wavelength of 195~$ {\buildrel _{\circ} \over {\mathrm{A}}}$, derived from ST-A/EUVI images.}
\label{fig:oneFIt}
\end{figure}
 
An example of an ecliptic J-map, showing two distinct families of converging tracks observed during July and August 2008, is presented in panel (a) of Figure~\ref{fig:oneFIt}. The convergent shape of each set of tracks is a characteristic of the corotating nature of the global structure \citep{2008GeoRL..3510110R}; the tracks converge along a `locus of enhanced visibility', indicated by green arrows \citep{2010ApJ...715..300S}. This locus marks the direction along the line-of-sight that is a tangent to the spiral formed by the series of emitted blobs, \textit{\textit{i.e.}} the CDS  \citep[see, \textit{\textit{e.g.}}][]{2010ApJ...715..300S}. To determine the time-dependent 3-D location of each CDS we select a single reference track in each pattern, corresponding to a single blob, and fit its trajectory by assuming that it moves radially outward at constant speed (the so-called `fixed-phi' approximation). It can be shown that the time variation of its elongation, $\alpha (t)$, can be described using the equation \citep[\textit{e.g.}][]{2010JGRA..115.4103R}:
\begin{equation}
\label{eq:fit}
\alpha(t)=\arctan \left[ { V_b t \sin \phi \over r_A(t) - V_b t \cos \phi} \right]  ,
\end{equation} 
where $V_b$ is the radial speed of the blob, $\phi$ is its propagation angle relative to the observer (which equates to ecliptic longitude relative to the observer for a feature propagating in the ecliptic plane\footnote{Positive $\phi$ angles, as seen from the ST-A point of view, mean that the propagation direction is east of the Sun--spacecraft line (this is the case throughout this work).}), and $r_A (t)$ is the radial distance of the observer from the Sun  (the observer is ST-A in this case). In the case of ST-B observations,  one simply needs only to replace $r_A (t)$ by $r_B (t)$. The elongation variation of the blob can be fitted using a two-parameter fit to retrieve estimates of $V_b$ and $\phi$. We use the same fitting procedure as described in Section 6 of \cite{2010JGRA..115.4103R}. Note that the radial motion of the satellite is also taken into account through the time dependence of its heliocentric radial distance $r_A$ (or equivalently $r_B$). It must be noted that the quality of the fit of one track is not the unique requirement for the overall corotating structure (\textit{i.e.} the CDS) to be represented correctly. In fact, fits to the tracks of several well-defined blobs are performed, and the one that provides the best representation of both the blob itself and the overall envelope is chosen as the reference blob for that CDS. \\ 

Once the trajectory of the most well defined small-scale transient (blob) is fitted for its radial speed and direction, we generate a pattern of converging tracks by using the following assumptions to model the entire CDS:

\begin{itemize}
 \item the best-fit radial speed of the reference blob is common to \emph{all} density inhomogeneities within the fitted CDS,
 \item the corotating structure is rotating with a fixed period of 25.38 days.
\end{itemize}

We ensure that the elongation variation of the tangent to the CDS (\textit{i.e.} the locus of enhanced visibility) maps closely the envelope of the converging tracks. In fact, the fit is only validated if other tracks are well fitted, and all tracks converge towards the locus of enhanced visibility.\\

Rearranging Equation~\ref{eq:fit} to express the blob speed in terms of $\alpha$ and $\phi$,
we can compute the minimal speed that a solar wind feature must 
have to cross two adjacent J-map bins during the 2-hour cadence. 
In this study we fit the apparent position of CDSs mostly inside
a 10$^{o}$--60$^{o}$ elongation window. We find that for typical
longitudinal separations of the radially-outflowing feature, $\phi$=10$^{o}$, 45$^{o}$, and 60$^{o}$, the minimal speed must be $\sim 30$ km~s$^{-1}$ or ten times 
less than typical solar wind speeds. At $\phi=$90$^{o}$, the minimal
speed is higher for larger elongations but at these large heliocentric distances the
blobs become undetectable. Overall, the time to cross two consecutive J-map bins is between
15 minutes and 1 hour at 300 km~s$^{-1}$ depending on the direction of propagation. 
The solar wind speeds we deal with in this work are never much less than 300 km~s$^{-1}$. Consequently, the J-maps offer sufficient precision to track solar wind features with a precision of $\sim 30$~km~s$^{-1}$ and to compare with
\textit{in situ} measurements with 1-hour cadence.\\

  An example of a fit to a single CDS, spanning July and August 2008, is shown in panel (b) of Figure~\ref{fig:oneFIt}. The red curve indicates the elongation variation of the fitted blob, emitted at 2008-08-03T21:36:26 UT\footnote{Emission time is defined as the time when the blob radial distance is equal to the solar radius.}. Its best-fit radial speed is $V_b =358 \pm 10$ km~s$^{-1}$ and it propagates at a best-fit longitude of $\phi = 34^{o} \pm 3^{o}$ relative to ST-A. The family of black curves in this panel are simulated (using Equation~\ref{eq:fit}), assuming regular release of a series of density structures at multiples of 8 hours relative to the release time of the blob associated with the red curve. These blobs cover the range of $\phi$ values decreasing from 180$^{o}$ to 0$^{o}$ with increasing time. The entire pattern comprises the CDS, and the elongation angle marking its outermost location at any one time (indicated by green arrows in panel a) is the time-varying location of the tangent to the overall spiral structure. This tangent is also indicated as a corresponding green arrow in panel (c), which shows the orbital configuration at the emission date 
 of the reference blob. ST-A is represented by the red spacecraft symbol, ST-B by the blue symbol and the Earth by the blue sphere between two spacecraft. The combined field of view of the HI cameras on ST-A is delimited by red lines. The blue spiral represents the corotating Parker spiral arm, propagating radially outward with the same speed as deduced from fitting the time-elongation profile of the blob corresponding to the red curve in panel (b). The red arrow represents the trajectory of that blob (34 degrees from the Sun -- ST-A line). Panel (d) shows the coronal map from the Carrington rotation 2073 at a wavelength of 195~${\buildrel _{\circ} \over {\mathrm{A}}}$, derived from ST-A/EUVI images. The back-projected source longitude of the CDS is at 352.2$^{o}$ Carrington longitude (indicated by a blue circle). A V-shaped coronal hole is clearly visible at 300$^{o}$--320$^{o}$ longitude, to the east of the source location. 

\subsection{Limitations of the use of J-maps for CDS identification}
\label{sect:limit_cases}
There are several factors that can affect the identification of CDSs in J-maps constructed from running-difference images. These include the following (as demonstrated later in the paper): 

\begin{itemize}
\item times when multiple CDSs pass through in the field of view at the same time. This can occur when the streamers are highly warped, for instance when the non-axisymmetric solar magnetic field becomes strong as solar activity increases;
\item times when multiple CMEs pass through the field of view at the same time as the passage of the CDS. This problem is important as solar maximum approaches (2011--2014). Given the high frequency of CMEs, it then becomes very difficult to identify CDSs in the images. We analyse this effect later in the paper;
\item times when the Milky Way passes through the field of view. This makes it extremely challenging to identify the passage of CDSs.
\end{itemize}

Consequently, there are a number of CDSs that are missed completely or misidentified when making such a catalogue derived from the ST-A heliospheric imagery.

\section{The first catalogue of CDSs derived from white-light imagery}

As part of the HELCATS project\footnote{ \hyperref[http://www.helcats-fp7.eu]{http://www.helcats-fp7.eu}}, the first catalogue of CDSs has been derived from ST-A heliospheric images from April 2007 (start of STEREO mission science phase, at solar minimum) up to August 2014 (start of reduced STEREO mission operations prior to superior conjunction, at solar maximum), based on criteria listed in the previous section. It is more challenging to derive the same catalogue from HI observations from ST-B for two reasons. Firstly, the quality of the images is not as good as for HI on ST-A because of instrumental issues \citep[\textit{e.g.}, ][]{2009SoPh..254..185B}. Secondly, the viewing direction of HI on ST-B off the west limb of the Sun (from the spacecraft perspective) changes the characteristic CDS pattern such that it consists of diverging tracks, as opposed to the convergent tracks observed from ST-A \citep{2010JGRA..115.4103R}. Hence, the presence of even one or two CME during the transit of a CDS through the HI field of view makes CDS identification in ST-B images far more difficult than for ST-A. Therefore at present the catalogue is restricted to the ST-A observations.\\

For each CDS, we record the back-projected emission time (at the solar surface) of the blob that produces the reference track in the ecliptic J-map. By fitting its trajectory, we also derive the heliocentric ecliptic longitudinal separation of this small-scale transient with respect to the ST-A spacecraft ($\phi$) and its average radial speed $V_b$. Because the fitting process also `folds in' a fit by eye to the entire pattern of converging tracks,  $V_b$ also represents the `average' radial speed of the entire CDS.  By back-projection, we also obtain an estimated source location for the fitted blob, from which, when combined with the launch time, we can assign to it a Carrington rotation number. Since we are analyzing a corotating structure, the source location of the fitted blob is the source location of all inhomogeneities constituting the CDS\footnote{For convenience we assigned a unique Carrington rotation number to the whole CDS instead of giving individual rotation numbers for each blob in the CDS, even though each has a different launch time.}. Using this estimate of the source location, we can use EUV images to identify the location of the closest equatorial coronal hole.  \\

\begin{figure}
\begin{center} 
 \includegraphics[width=0.95\textwidth]{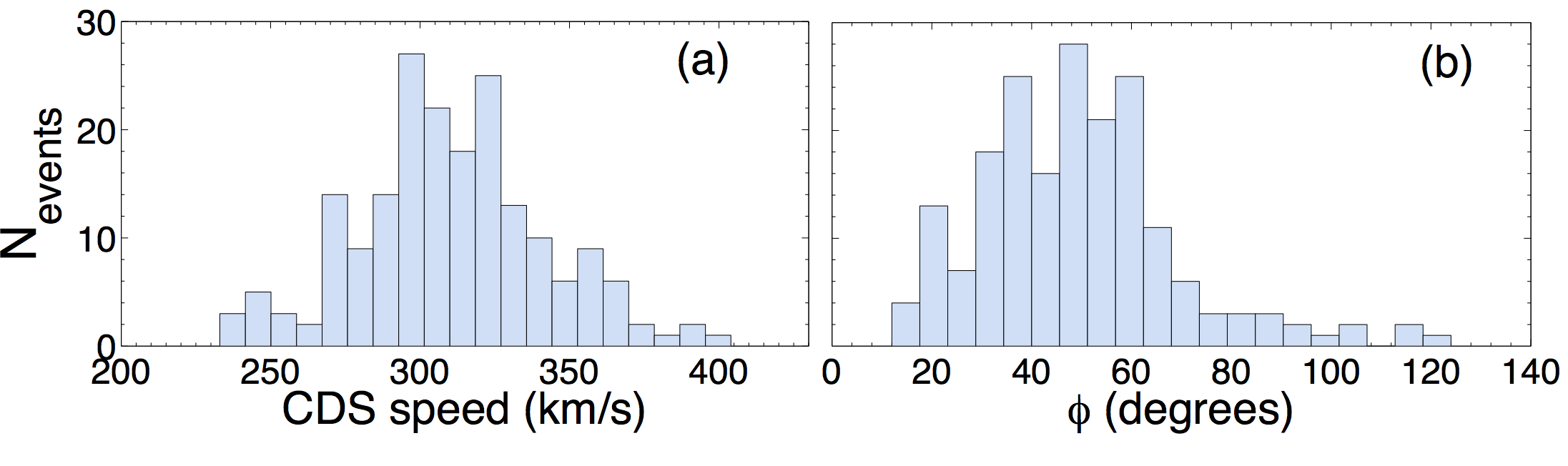}
\caption{Distributions of CDS speeds (panel a) and $\phi$ angles (panel b) for all catalogued events
observed between April 2007 and August 2014. These are derived by fitting the selected blob within each CDS. The mean speed is $\langle V_b \rangle = 311 \pm 31$~km~s$^{-1}$ and the mean direction is $\langle \phi \rangle = 49^{o} \pm 20^{o}$.}
\label{fig:histogram_SpeedBeta}
\end{center}
\end{figure}

\indent Figure~\ref{fig:histogram_SpeedBeta} presents histograms of the best-fit speed (a) and $\phi$ angle (b) for the selected blob tracked within each CDS. The mean speed $\langle V_b \rangle = 311 \pm 31$~km~s$^{-1}$ and mean angle $\langle \phi \rangle = 49^{o} \pm 20^{o}$. The maximum fitted speed is $404$~km~s$^{-1}$ and the minimum is $233$~km~s$^{-1}$. The range of speeds is roughly the same for both solar minimum (2007--2009) and solar maximum periods (2010-2014). No correlation was found between the level of solar activity and the average speed of CDSs. The Gaussian nature of the speed distribution has no physical basis \emph{a priori}. As can be seen from panel (b)  of Figure~\ref{fig:histogram_SpeedBeta}, the majority of fitted density blobs propagate at $\phi$ angles between 20$^{o}$ and 60$^{o}$. From the ST-A perspective, these separation angles  correspond to optimal visibility of the corotating spiral, because the line-of-sight in the ecliptic plane remains tangential to the spiral over a large range of elongations and density structures propagate close to the so-called Thomson sphere \citep[\textit{e.g.} ][]{2010ApJ...715..300S}. This explains why there are only a few events with $\phi>90^{o}$ and no events at all with $\phi>130^{o}$ while the maximum possible value is close to 180$^{o}$ (corresponding to a blob travelling away from the spacecraft). \\

\begin{figure}
\begin{center} 
   \includegraphics[width=0.8\textwidth]{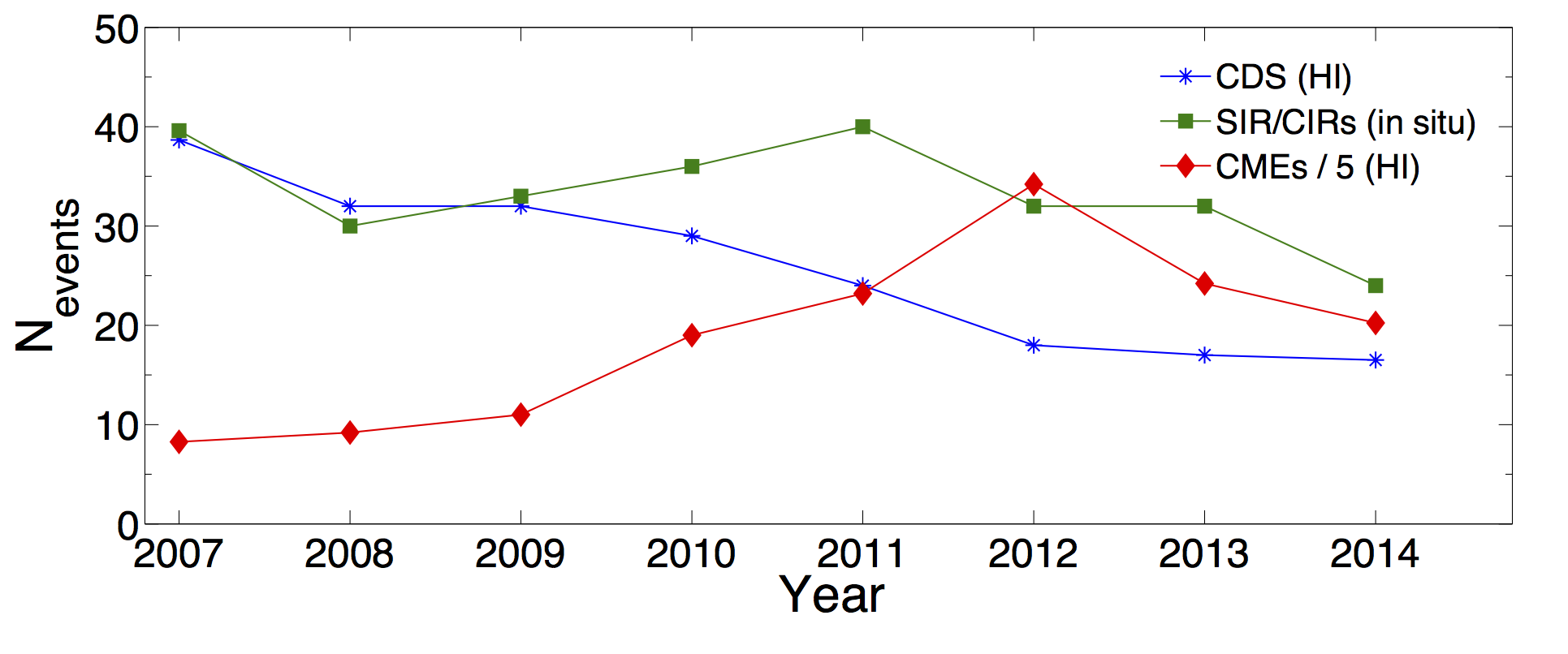}
  {\caption{Occurrence per year, from 2007 to 2014, of CDSs, SIRs/CIRs and CMEs. CDS occurrence, from our catalogue, is  plotted with blue stars, the occurrence of \textit{in situ} SIRs/CIRs, from L. Jian's  ST-A  catalogue \citep{2013AIPC.1539..191J}, is plotted with green squares, and red diamonds show the total number of CMEs detected in HI-1 images from ST-A (divided by 5 here for convenience). The latter was taken from the HELCATS website \hyperref[http://www.helcats-fp7.eu]{http://www.helcats-fp7.eu}. Incomplete years 2007 and 2014 are weighted accordingly.}}
\label{fig:Nevents}
\end{center}
\end{figure}

In this first statistical analysis of the solar origin of CDSs, we simply visually inspected Carrington maps constructed from 195~$ {\buildrel _{\circ} \over {\mathrm{A}}}$ EUVI images from both ST-A and ST-B for the presence of coronal holes, the westward edges of which are separated by less than 30$^{o}$ of longitude from the estimated source location of each CDS. It should be borne in mind that back-projecting radially to the solar surface, as we did, inherently ignores the complexity of the lower corona. The Carrington longitude of the directly back-projected coronal source region was catalogued for every CDS. For only 20 \% of CDSs was no coronal hole identified in the vicinity of their source location. The general presence of coronal holes to the east of the source location of ecliptic CDSs is strongly suggestive that such CDSs are mostly associated with the entrainment and compression of density inhomogeneities by high-speed (coronal-hole) streams. This is confirmed later in the current paper by the systematic comparison of the predicted arrival times of our identified CDSs with \textit{in situ} measurements. For those events associated with coronal holes, we also carried out the reverse analysis by locating the western boundary of the coronal hole and checking whether, for an average radial speed of $\sim$ 300 km~s$^{-1}$, we would predict the presence of an associated CDS in an ecliptic J-map. This was done for several events and gave a good agreement with the results from the analysis of the J-maps. \\

\indent  Figure~\ref{fig:Nevents} compares the solar cycle variation in the number of CDSs catalogued in the current study from  ecliptic J-maps and the number of SIR/CIRs detected \textit{in situ} at 1~AU from ST-A \cite[][]{2013AIPC.1539..191J}, as well as showing the number of CMEs identified in the HI-1 images from ST-A. The number of catalogued CDSs is above 30/year in 2007--2009 (\textit{i.e.} during solar minimum), roughly the same as the number of SIR/CIRs identified \textit{in situ} during the same period. The numbers of CDSs and SIR/CIRs diverge between 2010 and 2014, with a much lower number of CDSs (only 17 per year) being identified at solar maximum. One of the principal reasons for this drop in the occurrence rate of  CDSs is due to the difficulty in their identification, and accurate characterisation, in J-maps, during times when many CMEs pass through the HI field of view; at solar maximum a number of CDSs are ``hidden'' by the high number of CMEs crossing the field of view of the HI cameras (red curve). \\

\section{CDS propagation and validation versus \textit{in situ} measurements}

To determine the nature of a sample of the identified CDSs, we estimated their arrival times at a number of probes making \textit{in situ} measurements of the interplanetary plasma, namely ACE, \textit{Wind} and the STEREO spacecreft themselves. The cadence of \textit{in situ} plasma measurements we used was of 1 hour, being comparable to HI-2 images cadence (2~hours). This analysis was carried out for a subset of 61 events that occurred during 2007 and 2008. Focus was placed on the solar minimum period, when both the white-light images and \textit{in situ} data are easily interpretable due to the low occurrence rate of CMEs. The impact time at each \textit{in situ} probe of every CDS observed during that period was predicted and a comparison with the \textit{in situ} measurements undertaken. Since running-difference images reveal variations in plasma density, we compare our estimated CDS arrival times with the closest peak in density measured \textit{in situ}. The density peak corresponding to a SIR/CIR detected \textit{in situ} is usually located on the slow solar wind side of the stream interface, since the slow solar wind tends to be denser than the fast solar wind. Moreover, the highly dense heliospheric plasma sheet that is advected by the slow wind can find itself entrained and compressed by high-speed streams during their transit to 1~AU, enhancing the density asymmetry between the slow and fast solar wind plasma on either side of the stream interface. \\

The first question to address is whether the predicted passage of a CDS over an \textit{in situ} observatory occurs simultaneously with the observed passage of a SIR/CIR, the latter being a region of interaction between fast, tenuous, hot solar wind on one side and slow, dense, cold wind on the other. The fast and slow solar wind streams are separated by a stream interface where plasma density and transverse pressure maximise \citep[\textit{e.g.}][]{2006SoPh..239..337J}. The magnetic field, and the bulk solar wind ion speed, density and temperature, are used to identify the \textit{in situ} passage of a SIR/CIR. Typical SIR/CIRs are identified \textit{in situ} by the following signatures \citep{2010JGRA..11510101B}:

\begin{itemize}
\item a transition in the radial component of the solar wind speed from slow ($\sim$300 km~s$^{-1}$) to fast wind (500-700 km~s$^{-1}$).
\item a deflection in the flow around a stream interface.
\item a peak in plasma density very close to the stream interface, on the slow-wind side.
\item an amplification in the magnetic field in the compression region (\textit{i.e.} at the plasma density peak).
\item an enhancement in the following characteristic speeds: sound, Alfv\'{e}nic and magnetosonic speed, in the compression region.
\item a rapid increase in the ion specific entropy ($T_i/n^{2/3}$) at the stream interface.
\end{itemize}

\subsection{Comparison with \textit{in situ} measurements}
\begin{figure}
\begin{center}  
\includegraphics[width=\textwidth]{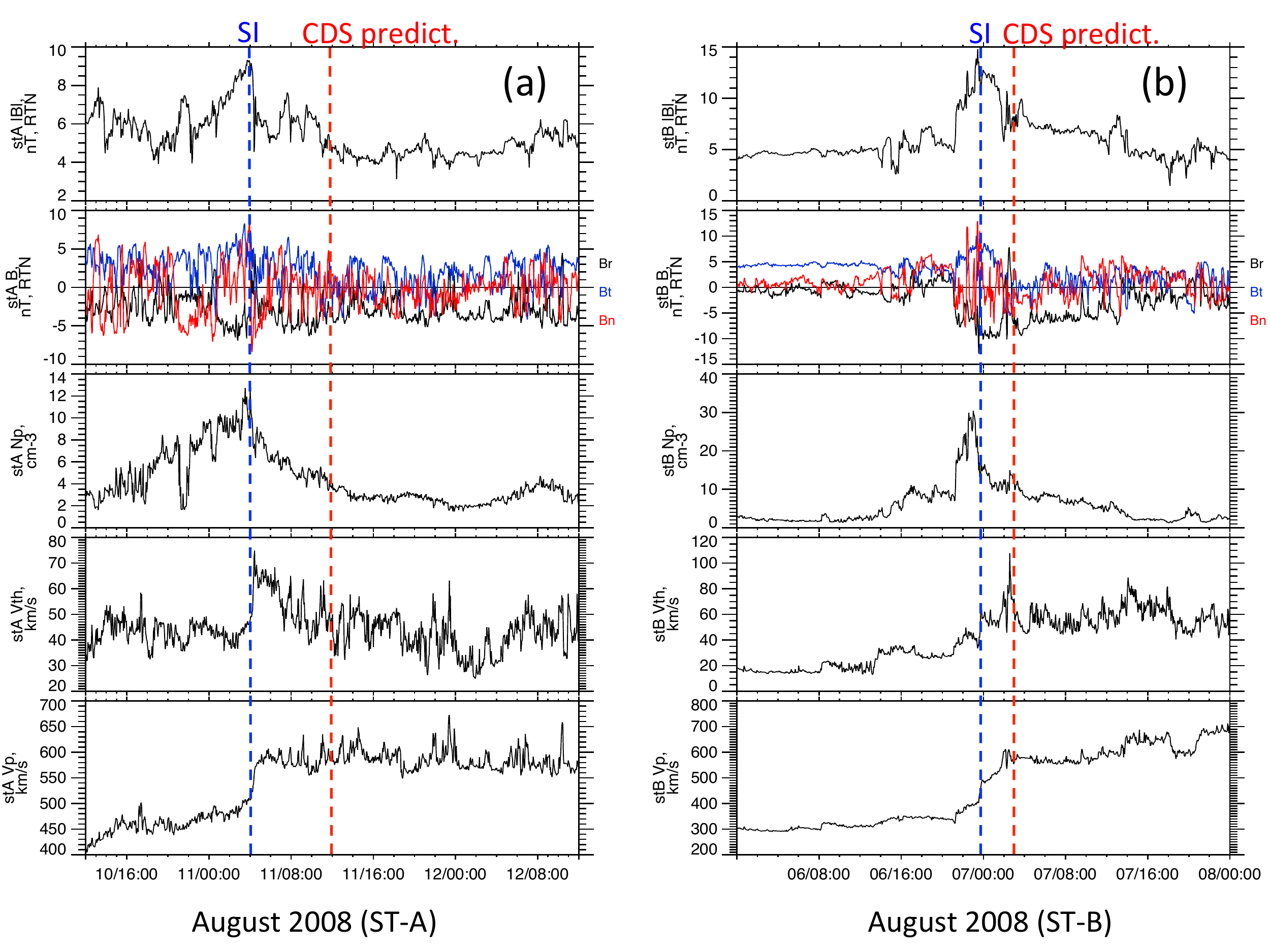}
\caption{\textit{In situ} data from ST-A (a) and ST-B (b) showing the SIR/CIR corresponding to the CDS fitted in Fig.\ref{fig:oneFIt}. From top to bottom: Magnetic field amplitude $|B|$, components of magnetic field $[B_r,B_t,B_n]$, proton density, thermal velocity, and bulk solar wind speed. The red vertical lines indicate the predicted arrival times of the CDS at ST-A and ST-B from the ST-A J-map analysis, which yields $V_b=358 \pm 10$~km~s$^{-1}$. The \textit{in situ} speed at the predicted time of arrival at both ST-A and ST-B is $\sim 600$ km~s$^{-1}$ (\textit{i.e.} in this case the predicted arrival time is too late, such that it lies on the fast wind side of the stream interaction region). Note that at the time of peak density (at the stream interface; indicated by the blue vertical lines), the solar wind speed is intermediate between the fast and slow wind speeds.}
\label{fig:insitu1}
\end{center}
\end{figure}

Figure \ref{fig:insitu1} presents the \textit{in situ} data taken by ST-A and ST-B spanning, in each case, the predicted arrival time of the CDS identified in Figiure~\ref{fig:oneFIt}. At both spacecraft, a SIR/CIR is clearly identified \textit{in situ}, close to the central time of each figure. SIR/CIR arrival is marked by an increase in magnetic field, a density enhancement, and a transition from cold, slow ($\sim 300$~km~s$^{-1}$),  dense plasma to hot, fast  ($\sim 600$~km~s$^{-1}$), tenuous plasma. The predicted arrival times at the two \textit{in situ} observatories, based on the analysis of the HI data from ST-A, are indicated by vertical red lines. Predicting the time at which the CDS would encounter ST-A and ST-B is done by adding, to the launch time of the fitted blob, the time required for its source region to corotate (at fixed solar rotation period of 25.38 days) to the longitude of the \textit{in situ} observatory and the time taken for such a feature to propagate (at its fitted speed) out to the heliocentric distance of that observatory. The mean predicted speed, $\langle V_b \rangle=358 \pm 10$~km~s$^{-1}$, is close to the slow solar wind speed measured at ST-A and ST-B prior to the arrival of the stream interface. In both cases, the predicted arrival time is slightly later than the \textit{in situ} detection time of the stream interface (by 6 hours for ST-A and 2 hours for ST-B). It should be noted that the method uncertainty in the predicted arrival time (corresponding to the uncertainty in the fit of the time--elongation profile of the reference blob) is of the order of $\pm 3$~hours if the uncertainty in the mean speed is propagated over 1~AU.  An analogous comparison was done using ACE and \textit{Wind} data. For this particular CDS, the predicted arrival time is several hours after the \textit{in situ} detection of the stream interface at all of the probes considered. \\

\begin{figure}
\begin{center} 
\includegraphics[width=\textwidth]{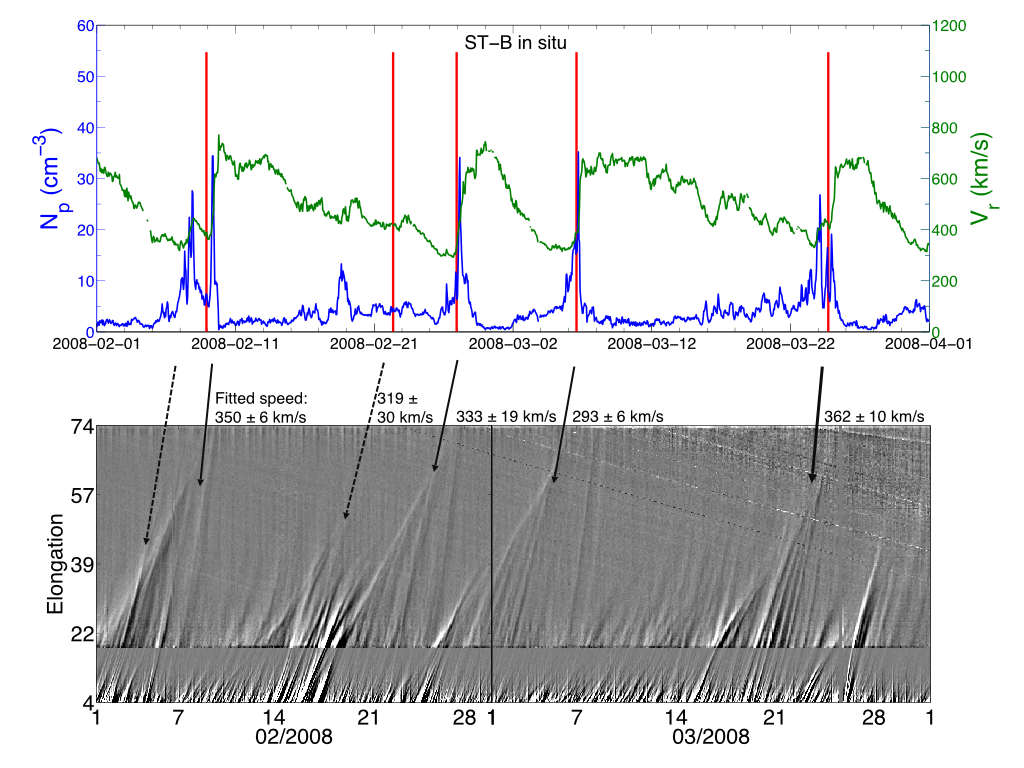}
\caption{Illustration of a series of  SIR/CIRs detected \textit{in situ} at ST-B (upper panel) and their corresponding signatures in ST-A HI J-maps (bottom image). The blue
and red time series presented in the upper panel correspond to the density $N_p$ and the solar wind speed $V_r$. Vertical red lines indicate the predicted arrival times at ST-B of the five CDSs shown in the ST-A J-maps (in the bottom panels). Black arrows indicate HI signatures of the \textit{in situ} CDS predicted events. The first black arrow does not correspond to a CDS but preasumably to a CME.}
\label{fig:insitu2}
\end{center}
\end{figure}

We present, in the upper panel of Figure~\ref{fig:insitu2}, the \textit{in situ} measurements made by ST-B around the predicted impact times of five CDSs with the arrival time of each CDS being marked by a red vertical line. The proton density and the radial solar wind speed are plotted over a time window extending from 1 February 2008 to 31 March 2008. The bottom panel of the figure shows ST-A J-map with  signatures  of theses five different CDSs, which correspond to four different SIR/CIRs detected \textit{in situ} during this time period (top panel). The second CDS (indicated by the second dashed arrow) has no SIR/CIR counterpart \textit{in situ} at ST-B. Each SIR/CIR is followed by a long-lasting high-speed stream. Before the arrival of the first SIR/CIR (02-02 to 10-02), multiple magnetic flux ropes are detected \textit{in situ}, each of which is associated with a CME clearly identified near the Sun (first dashed arrow). The associated J-map is complex, with multiple CME tracks crossing the (albeit still clear) CDS signature. The \textit{in situ} signature suggested to be associated with the second CDS is the most interesting, as it does not correspond to that of a typical SIR/CIR or even have a strong density peak at ST-B (ST-B was situated, at this time, at a  heliographic latitude of --5.6$^{o}$). For this event, we find in the heliospheric imagery that the compression by the high-speed stream mainly occurs south of the ecliptic plane; the weak CDS detected in the ecliptic by HI is likely the signature of the northern boundary of the SIR/CIR. The associated SIR/CIR is, however, detected by ST-A \textit{in situ} (at a heliographic latitude of --7.2$^{o}$). Because the ecliptic transients associated with this event experience no significant compression during their outward propagation, their brightness decreases rapidly as they propagate outward and their J-map signature is weak. The third and fourth CDSs have typical SIR/CIR signatures \textit{in situ} \citep[see \textit{e.g.}, ][]{2010JGRA..11510101B}.\\ 

\subsection{Long-term variations in the properties of CDSs}
\label{s:long_term}
\begin{figure}
\begin{center}
\includegraphics[width=\textwidth]{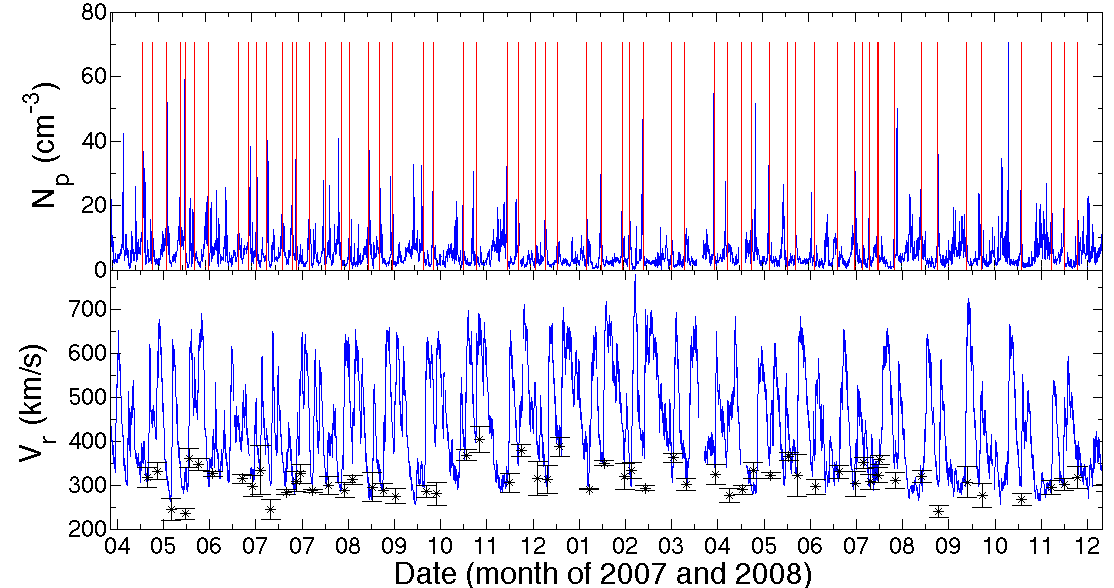}
\caption{ST-A \textit{in situ} measurements of the proton density (top panel) and the radial solar wind speed (bottom) from April 2007 to December 2008. Red vertical lines indicate the predicted CDS arrival times. 
Black stars in the lower panel, with corresponding error bars, are the predicted CDS speeds from J-map analysis, plotted at the predicted arrival times. Quiescent density values do not exceed 10 cm$^{-3}$ while strong peaks reveal the presence of local strong compression regions. The speed values oscillate between 300 km~s$^{-1}$ (slow solar wind) and 700 km~s$^{-1}$ (fast wind).}
\label{fig:global_ST-A}
\end{center}
\end{figure}

To illustrate the connection between CDSs and SIR/CIRs during solar minimum,  Figure~\ref{fig:global_ST-A} presents ST-A measurements of the proton density (upper panel) and the radial solar wind speed (lower panel) from April 2007 to December 2008. This period corresponds to the deep solar minimum, with very few CMEs emerging from the corona. The recurrent pattern of fast ($>$600 km~s$^{-1}$) then slow solar wind ($<$400 km~s$^{-1}$) is clearly visible in the lower panel. Very few of the large density increases are not associated with a predicted CDS impact. For these events, we find that either the J-map was of insufficient quality to permit the definitive identification of a CDS or that the density peak measured \textit{in situ} was associated with the passage of a CME rather than a SIR/CIR. Red vertical lines in the upper panel  (density) indicate the predicted impact times of all CDSs catalogued during this interval. The black stars  in the lower panel show the fitted CDS speeds ($V_b$) at the predicted impact times. The fitted speeds are clearly close to the slow solar wind speed. Remarkably, the time variation of the fitted speeds follow closely the time variation in the speed of  the slow wind ahead of the SIR/CIRs detected \textit{in situ}. For these reasons, we conclude that these CDSs, at least, are mainly associated with the outward propagation of SIR/CIRs and with the formation of strong density enhancements in the interplanetary medium. Moreover, during their transit from the Sun to 1~AU, the CDSs appear to propagate with a speed close to that of the slow solar wind rather than the typical average speed of stream interfaces located inside SIR/CIRs (which tends to be some 100~km~s$^{-1}$ faster than the slow solar wind speed).  \\

\begin{center}  
 \begin{figure}
 \includegraphics[width=0.99\textwidth,clip=]{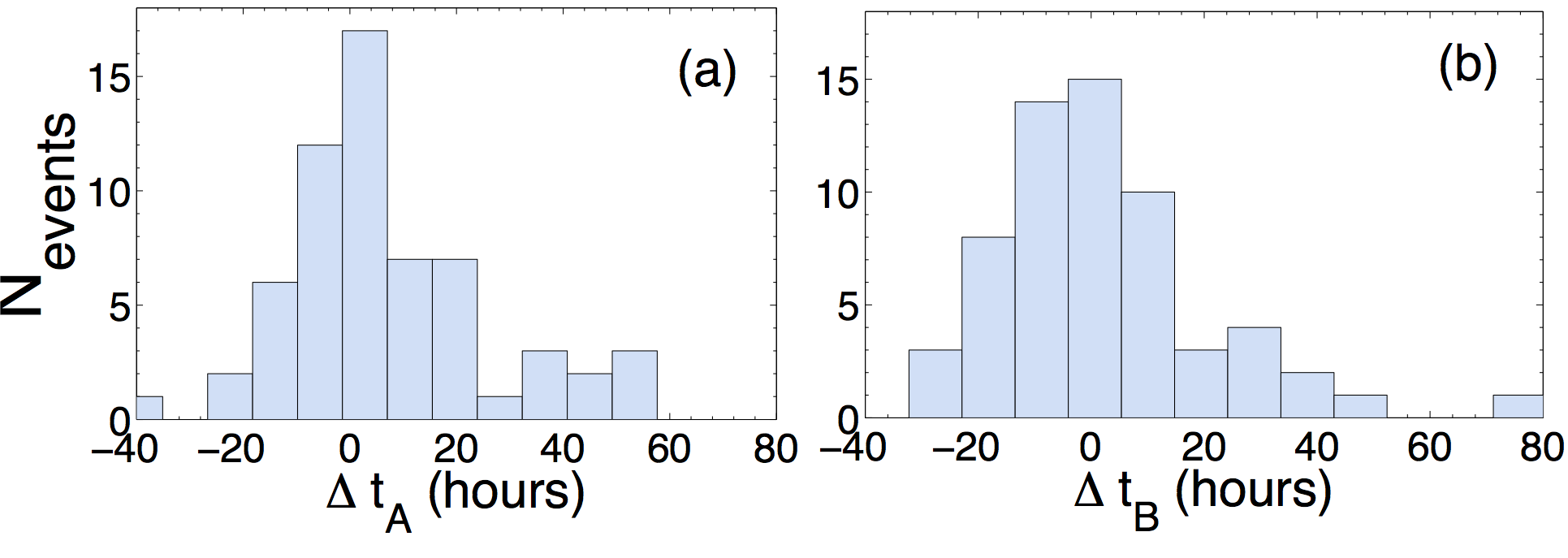}
\caption{Histograms of differences between the predicted CDS arrival time and the time of the density peak detected \textit{in situ} $\Delta t = t_{pred} - t_{peak}$ for ST-A (a) and ST-B (b). All events catalogued during  2007 and 2008 were used to generate these plots. This period corresponds to the deep solar minimum.}
\label{fig:time_Delays}
\end{figure}
\end{center}

A closer look at Figure~\ref{fig:global_ST-A} shows that the predicted CDS arrival times do not coincide exactly with times of the SIR/CIR density peaks detected \textit{in situ}. The time of passage of the \textit{in situ} density peak of the SIR/CIR associated with each CDS (this association was done by eye) was recorded for every event, as was the corresponding difference between the predicted and measured arrival times ($\Delta t = t_{pred} - t_{peak}$).  Figure~\ref{fig:time_Delays} presents histograms of $\Delta t$  for ST-A (panel (a)) and ST-B (panel (b)). Statistically there is a tendency for the predicted impact time to be later than the \textit{in situ} arrival time of the density peak (corresponding to positive values of $\Delta t$). This is consistent with our observation that the CDS speed tends to be lower than the \textit{in situ} speed of the SIR/CIR. The mean values of the time differences for ST-A and ST-B (denoted by $\langle \Delta t_A \rangle$ and $\langle \Delta t_B \rangle$) are $+6.5$ hours and $+2.4$ hours, respectively; standard deviations are of the order of 19 hours in each case. Note that the minimum absolute value of $|\Delta t|$ is 2 minutes and the maximum value is 3 days. The most probable absolute values of $\Delta t_A$ and $\Delta t_B$ are $9$~hours and $11$~hours, respectively. Assuming that the density peak corresponds to the location of the stream interface, $\Delta t > 0$ will tend to mean that the arrival time of the CDS is predicted to lie in the faster solar wind regime downstream of the interaction region while $\Delta t < 0$ will tend to mean that the CDS is predicted to arrive in the slower solar wind regime upstream of the stream interface.

\begin{figure}
\begin{center}
 \includegraphics[width=\textwidth,clip=]{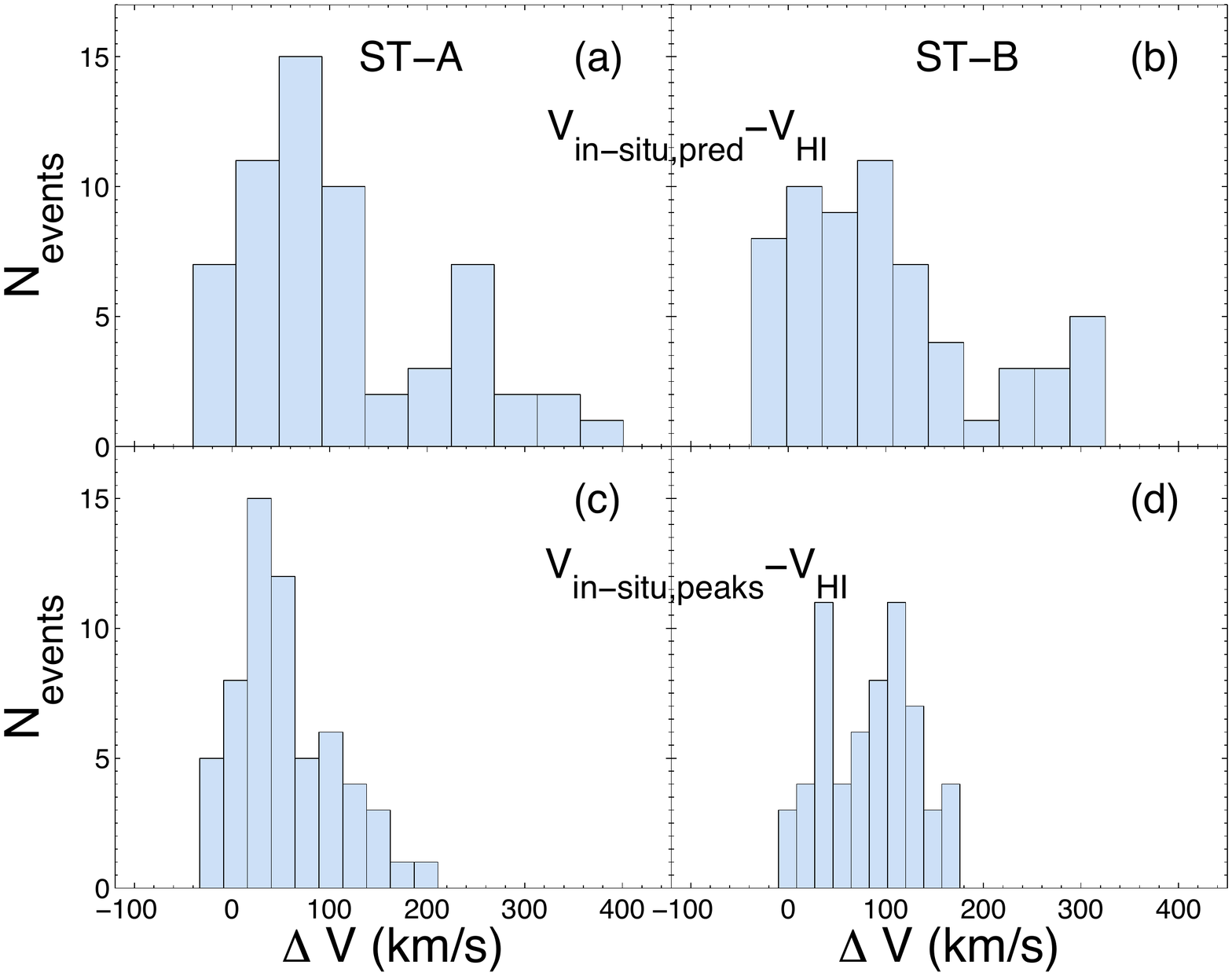}
\caption{Histograms of differences between \textit{in situ} radial solar wind speed measured at the predicted CDS impact times and the CDS radial speed from our ST-A catalogue (a,b), and between \textit{in situ} radial solar wind speed measured at the time of the associated SIR/CIR density peak and the CDS radial speed from our ST-A catalogue (c,d). Left and right panels are based on ST-A and ST-B \textit{in situ} measurements, respectively.}
\label{fig:compare_speed}
\end{center}
\end{figure}

In Figure~\ref{fig:compare_speed} we compare the distribution of the \textit{in situ} radial solar wind speed at the predicted CDS impact times with that of CDSs speed (panels (a) and (b)). We also plot the distribution of the differences between the \textit{in situ} radial solar wind speed measured at the time of the density peak associated with (\textit{i.e.} nearest in time to) each predicted CDS arrival and the CDS speed (panels (c) and (d)). CDS speeds, derived from the analysis of the ST-A J-maps, form a distribution centered at $311$~km~s$^{-1}$ with a half-width of $30$~km~s$^{-1}$. The distribution of \textit{in situ} speeds measured at ST-A at the predicted CDS arrival times spans a range of speeds extending from $300$~km~s$^{-1}$ (slow wind) to $>600$~km~s$^{-1}$ (fast wind). The same is true for  ST-B. The broad nature of this distribution, extending up to the speed of the fast solar wind ($\Delta V \geq 100$), is a consequence of the distribution in arrival time errors, $\Delta t$,  shown in Figure \ref{fig:time_Delays}. The relatively large errors in the predicted CDS impact time relative to the stream interface move it into the slow wind regime for $\Delta t<0$ (upstream of the stream interface) and into the fast wind for $\Delta t >0$ (downstream of the stream interface). This distribution shows a tendency to measure a significantly faster \textit{in situ} solar wind than the CDS predicted speed, due to aforementioned tendency for a larger part of the predicted CDS arrival time to be later, even, than the arrival of the density peak and hence in the fast solar wind regime. As illustrated in panels (c) and (d) of Figure~\ref{fig:compare_speed}, when the times of the \textit{in situ} density peaks are considered instead, the distribution of $\Delta V$ becomes narrower, centered around 50 km~s$^{-1}$ for ST-A  and 100 km~s$^{-1}$ for ST-B. There are no more events exceeding $\Delta V = 200$~km~s$^{-1}$.\\

 \subsection{Comparison with \textit{in situ} catalogues of SIR/CIRs}
 \label{subsect:insitu_compar}
 
\begin{figure}
 \begin{center}
\includegraphics[width=0.75\textwidth]{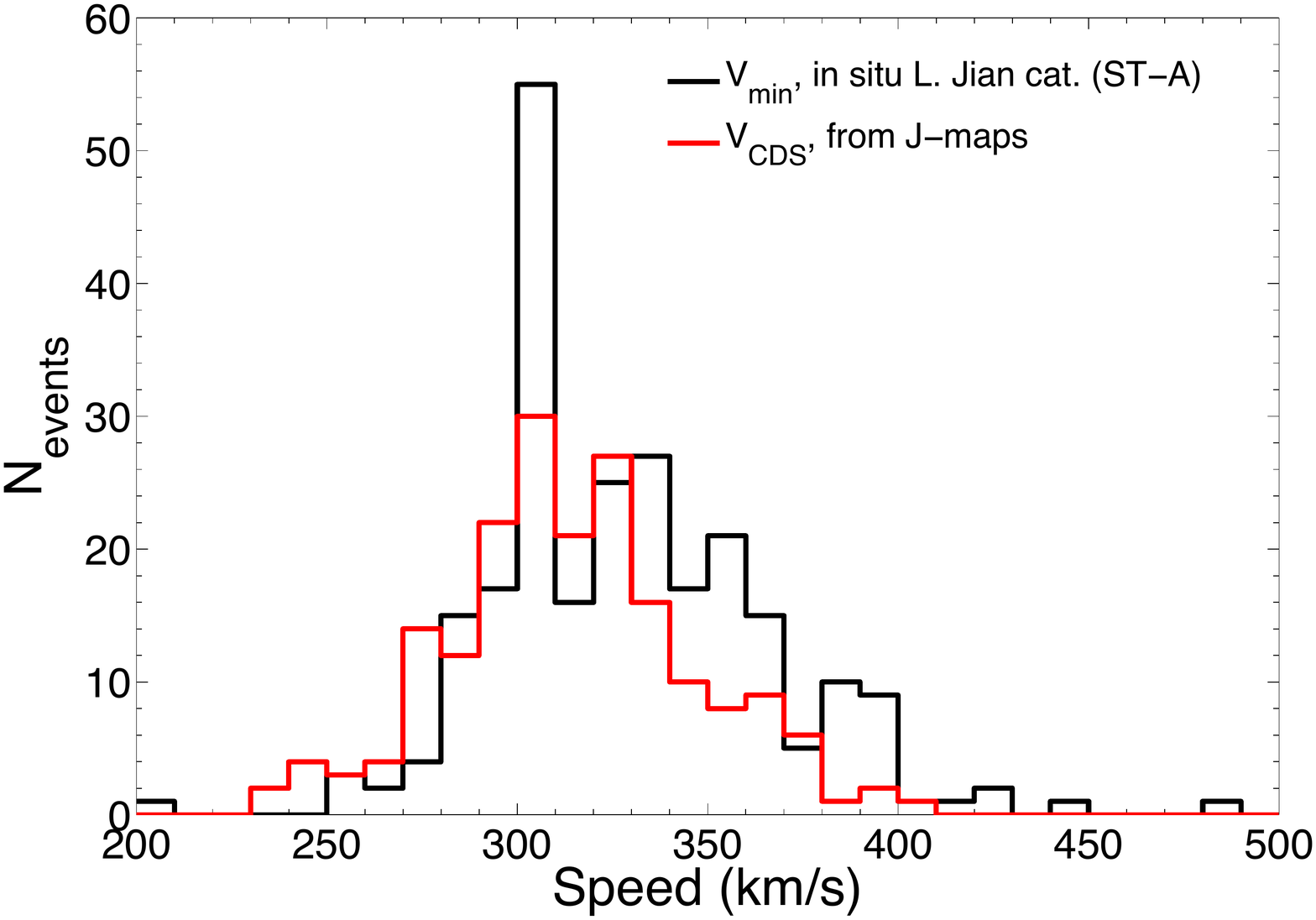}
\caption{Distributions of the derived CDS speed (red histogram) and of the minimum solar wind speed (black histogram) prior to the arrival of the stream interface at ST-A. The latter are taken from the \textit{in situ} catalogue of SIR/CIRs derived by \cite{2013AIPC.1539..191J}, but only for ST-A.}
\label{fig:jian_compare}
\end{center}
\end{figure}

 \textit{In situ} catalogues of SIR/CIRs have been produced using ACE and \textit{Wind} datasets for the interval spanning the years 1995 to 2009  \citep{ 2006SoPh..239..337J,2011SoPh..274..321J} and using STEREO datasets for the years 2007 to 2014 \citep[][catalogue updated online]{2013AIPC.1539..191J}. These catalogues provide a comprehensive list of SIR/CIRs and their physical properties (including the solar wind speed before and after the stream interface, as well as the maximum values of the density, pressure and magnetic field). The information contained within the STEREO SIR/CIR catalogue (which includes 248 events from ST-A and 231 from ST-B) is compared here with our predictions.  \\
 \indent The first point to note is that we see fewer events in total (190) than are detected \textit{in situ} over the same period. As shown in Figure~\ref{fig:Nevents}, over the solar minimum period (2007--2009) roughly the same number of events were imaged in HI as were detected \textit{in situ} (30--40/year). At solar maximum far fewer CDSs were imaged by HI (17/year) while the frequency of SIR/CIRs detected \textit{in situ} remained almost unchanged. This discrepancy is, as has been discussed at length above, attributed to the increasing number of CMEs at solar maximum. \\
 \indent Next, we compare the CDS speeds with the minimum solar wind speed measured ahead of the SIR/CIR compression region and also with the speed at the stream interface itself. To this end, Figure~\ref{fig:jian_compare} displays the distributions of the speeds of our 190 catalogued CDSs (in red) and of the minimum speed of the slow solar wind measured ahead of the SIR/CIRs, detected by ST-A \citep[black; ][]{2013AIPC.1539..191J}. Generally, the two distributions seem to be in good agreement, with the mean CDS speed of 311 km~s$^{-1}$ being comparable with the mean minimum slow solar wind speed of 325 km~s$^{-1}$ (expressed as $V_{min}$ in the catalogue of \cite{2013AIPC.1539..191J}\footnote{320 km~s$^{-1}$ at ST-B}). In contrast, the mean speed at the stream interface itself, the latter identified as the location of maximum transverse pressure $P_t$ in \cite{2006SoPh..239..337J} is 424 km~s$^{-1}$ at ST-A and 404 km~s$^{-1}$ at ST-B. These latter values lie close to 411 km~s$^{-1}$, which corresponds to the average speed of the \textit{in situ} density peak closest to the predicted arrival time (see Figure~\ref{fig:compare_speed}). This provide further hints that (1) the CDS speeds are globally those of the slow solar wind just ahead the stream interface at 1~AU and (2) the closest density peaks to the predicted impact times lie very near the stream interface.  \\

 \section{ST-A CDS catalogue from 2007 to 2014}

 \begin{figure}
 \begin{center}
\includegraphics[width=\textwidth]{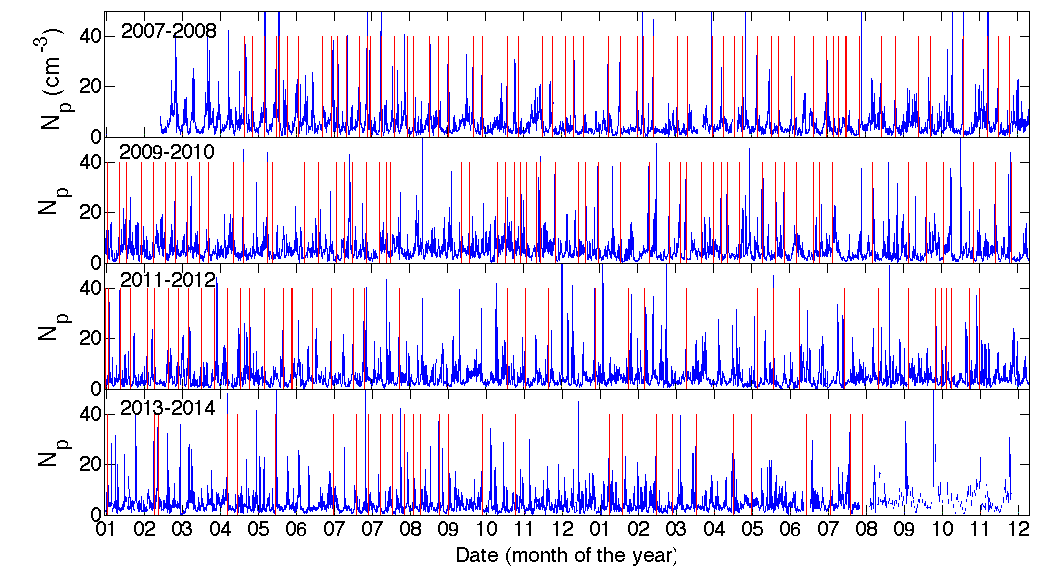}\\
\includegraphics[width=\textwidth]{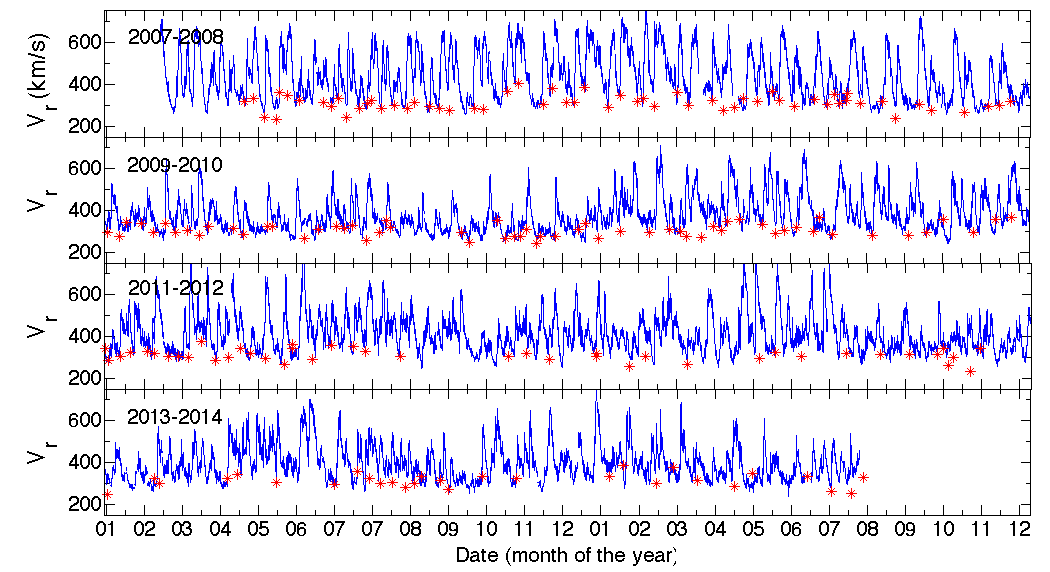}
\caption{\textit{In situ} proton density (top) and radial solar wind proton speed (bottom) as measured by the ST-A spacecraft from 2007 to 2014. Over-plotted are predicted impact times of all catalogued CDSs at ST-A (top figure: red vertical bars) and fitted CDS speeds (bottom: red stars).}
\label{fig:global_allYears}
 \end{center}
\end{figure}

 Here we summarize the catalogue of CDS events and the results of their propagation to a number of \textit{in situ} observatories. The ST-A HI CDS catalogue from April 2007 to August 2014 includes 190 events \footnote{Catalogue content available at the HELCATS website: \hyperref[http://www.helcats-fp7.eu/catalogues/wp5_cat.html]{http://www.helcats-fp7.eu/catalogues/wp5\_cat.html}}. For each CDS, the information provided in this catalogue is: the launch time of the fitted blob ; its speed $V_b$ ; its direction $\phi$ relative to ST-A ; the Carrington rotation number of the CDS; the Carrington longitude of the CDS source location; the predicted impact times at different probes near 1~AU (ST-A, ST-B, \textit{Wind} and ACE). The catalogue covers the ascending phase of Solar Cycle 24, from solar minimum (2007-2009) to solar maximum ($\sim$2010-2014). As was illustrated in Figure~\ref{fig:Nevents}, the number of well identified CDS events detected by HI on ST-A appears to be anti-correlated with the number of CMEs observed by HI. As the number of CDSs falls, from $\sim$40/year in 2007 to 17/year in 2014, the number of CMEs rises from 40/year in 2007 to a maximum of several hundred/year in 2012, after which it decreases slightly. CME activity prevents the systematic identification of corotating structures in HI imagery. \\
 \indent Figure~\ref{fig:global_allYears} presents the density (top) and radial speed (bottom) of the solar wind measured \textit{in situ} at the ST-A spacecraft, from 2007 to 2014. Predicted arrival times (red vertical lines: top panel) and speeds (red stars: bottom panel) of all catalogued CDSs are over-plotted. During the first half of the period under consideration (2007--2010), density structures that are not associated with predicted CDS arrival are rare, and the fitted speeds follow well the slow solar wind variations. Note that the uncertainty in the arrival time at 1~AU is about 11 hours (Section~\ref{s:long_term}). By contrast, between 2011 and 2014 the number of fitted CDSs decreases markedly. This is because of the difficulty of observing ``clean'' CDS signatures (a well constrained fit requires the clear identification of the tracks of several individual blob tracks, as well as a clear envelop) due to high level of CME activity (see, \textit{e.g.} Section~\ref{sect:limit_cases}). 

\section{The nature of CDSs}
 \begin{figure}
 \begin{center}
\includegraphics[width=0.94\textwidth]{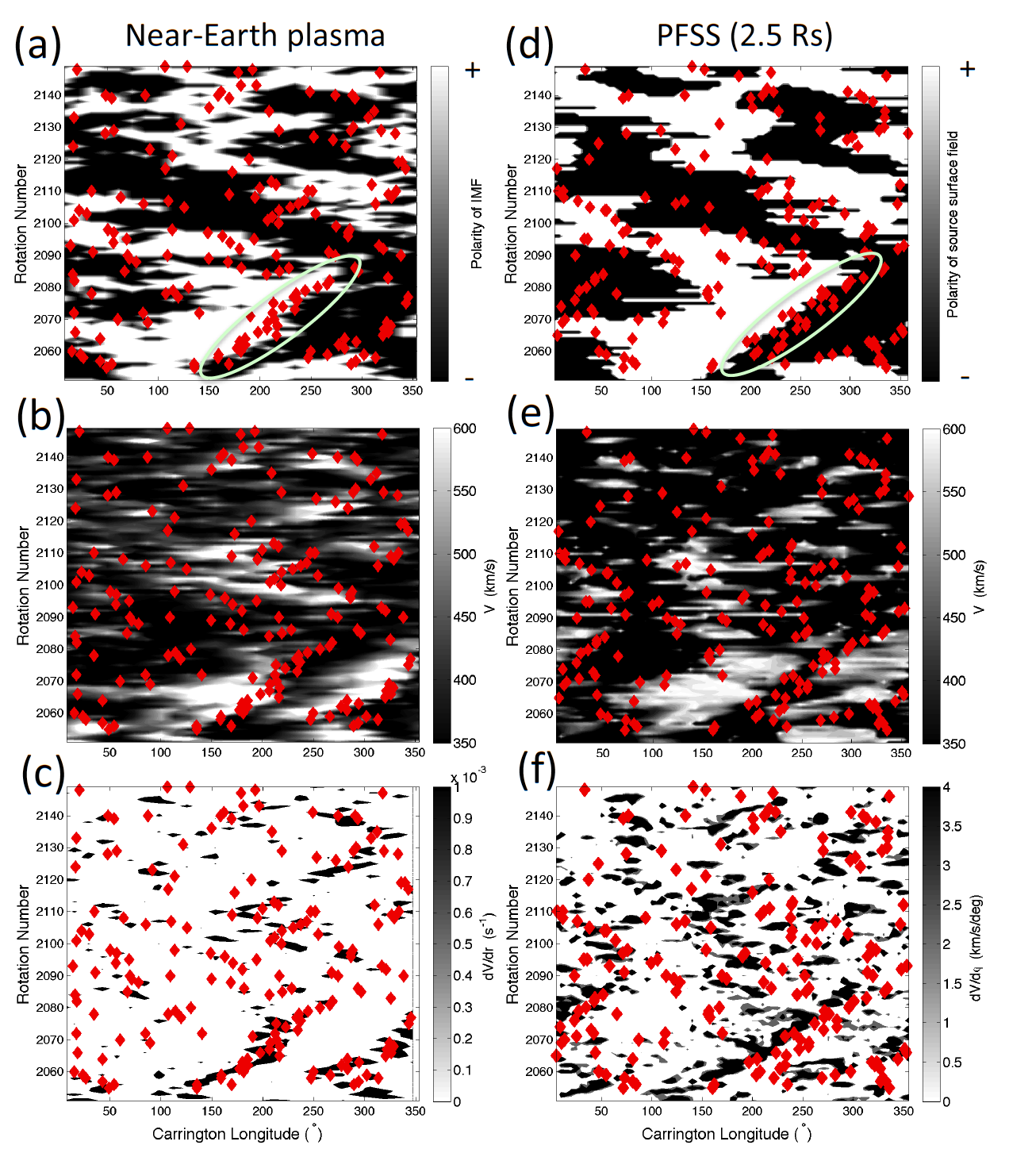}
\caption{A series of contour plots of \textit{in situ} measurements of the near-Earth (L~1) plasma (panels a-c) and of the solar magnetic field from a PFSS model (panels d-f). All panels are presented in the same format, with Carrington longitude running along the X-axis and Carrington rotation number along the Y-axis. The red diamonds indicate the estimated Carrington rotation number versus longitude of each CDS in our catalogue. Note that time runs from high to low values of Carrington longitude (right to left along the X-axis) and low to high Carrington rotation number (bottom to top along the Y-axis). }
\label{fig:stackplot}
 \end{center}
\end{figure}

To analyse the origin and evolution of CDSs over the solar cycle, and interpret the observations shown earlier, we present a series of contour plots in Figure~\ref{fig:stackplot} of \textit{in situ} measurements of the near-Earth (L~1) plasma (panels (a)--(c)) and of the solar magnetic field derived using a PFSS model \citep[panels (e)--(g); ][]{1992ApJ...392..310W}. All contour plots are presented in the same format, with Carrington longitudes running along the X-axis and Carrington rotation number along the Y-axis. Carrington rotation numbers range from 2051 to 2150 \textit{i.e.} from January 2007 to May 2014. We remind the reader that, on these stack plots, time runs from high to low values of Carrington longitude (right to left along the X-axis) and from low to high numbers of Carrington rotation number (bottom to top along the Y-axis).  All measurements correspond to magneto-plasma parameters (described below) either directly measured in (left-hand columns) or extrapolated to (right-hand column) the ecliptic plane. The red diamonds overlaid in each panel show the estimated Carrington rotation number versus longitude of each CDS in our catalogue. \\
\indent Panel (a) shows the polarity of the interplanetary magnetic field derived from the OMNI dataset \citep{2005JGRA..110.2104K}, converted from RTN coordinates to azimuth angle, $\phi_{IM}$. We consider that all field orientations that are within 90$^\circ$ of the average Parker spiral orientation azimuths of 45$^\circ$ and 315$^\circ$ have negative and positive polarity, respectively. We have shifted the times of \textit{in situ} measurements backward by 5 days to roughly account for propagation time to 1~AU. In this way we have associated to the resulting `solar date' of the measurement, a Carrington rotation number and a longitude. The resultant data, displayed in panel (a), demonstrates the well-known sector structure of the interplanetary magnetic field as well as showing the transitions between polarities that mark the passages of the heliospheric current sheet. This panel shows that a magnetic field sector structure comprising between two and four sectors existed in the ecliptic plane throughout this weak solar cycle. The red diamonds correspond to the predicted Carrington rotation number versus longitude of all catalogued CDSs (\textit{i.e.} the results from the previous section). Examination of the locations of these red diamonds show that CDSs occur most frequently at locations where the interplanetary magnetic field polarity reverses, \textit{i.e.} at the heliospheric current sheet; this is particularly clear during the solar minimum years (see the region bounded by the green oval) but is also visible at solar maximum.\\
\indent Panel (b) enables comparison of the CDS arrival time with the radial solar wind speed measured \textit{in situ}. The red diamonds tend to cluster in those regions where the solar wind speed increases (\textit{i.e.} changes from black to white) rather than decreases  (changes from white to black). This is more clearly seen in panel (c), which presents the radial gradient in the solar wind speed. Panels (b) and (c) confirm the association between CDSs and SIR/CIRs demonstrated in the previous section. Interestingly, there are rare cases of CDSs occurring on the rarefaction side of coronal holes, where the solar wind speed decreases with decreasing Carrington longitude (\textit{e.g.} between $100$ and $120^\circ$ longitude, between rotations 2070 and 2080). These occur, nevertheless, on a magnetic sector boundary. We will discuss these events further in the next section. \\
\indent  Since we include the  Carrington longitude and estimated launch time of each CDS in the catalogue, we can also compare these parameters with reconstructions of the coronal magnetic field. We apply the potential field source surface (PFSS) technique of \cite{1992ApJ...392..310W} to magnetograms taken by the Wilcox Solar Observatory (WSO). Panel (d) shows the resultant polarity of those magnetic field lines threading the source surface set at 2.5 $R_\odot$ that are connected to the Carrington coordinates of Earth. The PFSS-derived sector structure is in very good agreement with the sector structure derived from \textit{in situ} measurements. Again there is a clear tendency for CDSs to cluster around the neutral line (\textit{e.g.} region bounded by the green oval). This provides further evidence for a close association between CDS and polarity inversions near the Sun. Nevertheless, there is evidence, both in the panels presenting the \textit{in situ} data and the PFSS output, of the rare occurrence of CDSs occurring far from the neutral line; these could be investigated in future studies.  \\
\indent For completeness, we also predicted the speed of the solar wind streaming out of flux tubes reconstructed using the PFSS model. To do this, we computed the flux tube expansion factor $F_S$, the amount by which the magnetic flux tube expands in solid angle between the photosphere and the source surface \citep{1990ApJ...355..726W}, for each flux tube connected to the Earth. \cite{1990ApJ...355..726W} showed that the expansion factor  is anti-correlated with the solar wind speed, $V$. To perform this analysis, we used the following relationship between the expansion factor and the solar wind speed, which has been shown to reproduce roughly the solar wind speed measured near 1~AU \citep{2007JGRA..112.5103R}: 

\begin{equation} 
V=280+350\times e^{-(F_S/14)^2} .
\end{equation} 

Panel (e) of Figure~\ref{fig:stackplot} displays the predicted solar wind speed at the source surface ($2.5~R_\odot$) and panel (f), the longitudinal gradient $-dV/d\phi$ in the predicted solar wind speed  (the negative sign in the latter expression being related to the decrease in Carrington longitudes with time).  This longitudinal gradient will be manifested, as the Sun rotates, in a radial solar wind gradient in the interplanetary medium (\textit{e.g.} panel (c)). The red diamonds cluster predominantly in slow solar wind regions (panel (e)), with a significant fraction lying close to the boundary with rising solar wind speed. Just as for panels (b) and (c), there are cases of CDSs originating within the predicted source regions of high speed solar wind but still near the sector boundary (as shown in panel (a)).\\

 \begin{figure}
 \begin{center}
\includegraphics[width=0.67\textwidth]{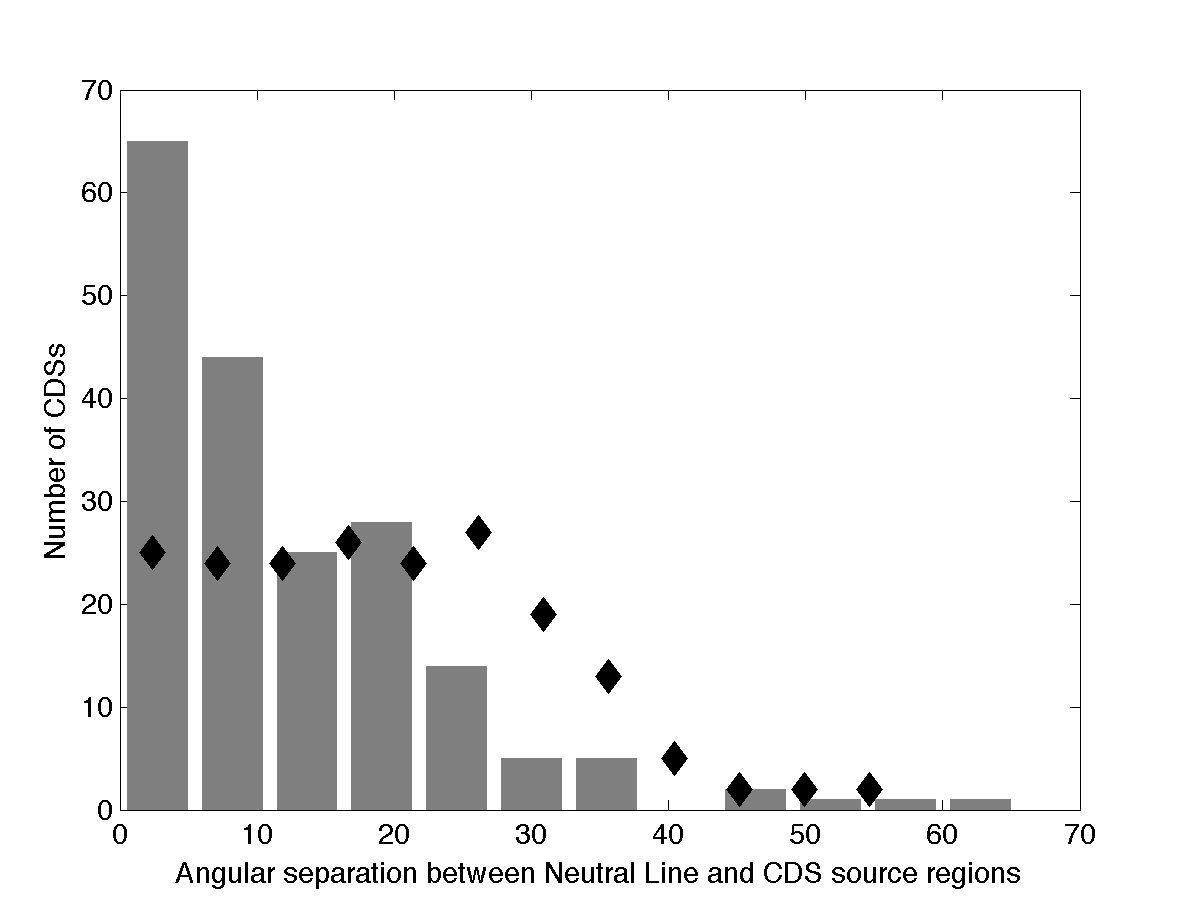}
\caption{Distribution of the angular separation, $\Delta \gamma$, between the PFSS neutral line and the recorded CDS source location at the Sun. The distribution resulting from random source locations (instead of fitted CDS source locations) is plotted using black diamonds. }
\label{fig:histoneutral}
 \end{center}
\end{figure}

The clearest relationship seen in the right-hand contour plots is between CDS location and the magnetic sector boundary. To test this correspondence more quantitatively, we show (as a histogram in Figure~\ref{fig:histoneutral}) the distribution of the angular separation, $\Delta \gamma$, between the PFSS neutral line and the recorded CDS source locations at the Sun. Half of the 190 catalogued in-ecliptic CDS events are separated by less than 10$^{o}$ from the neutral line. However, most of our catalogued CDSs occurred at solar minimum, when the neutral line tends to be less warped than at solar maximum, and therefore when the separation between sources in the ecliptic plane and the neutral line would be expected to be small. Hence, in order to test the significance of this result we compared the distribution of observed  $\Delta \gamma$ values with one that would result from a random distribution of the same number of source locations. We find that the peak in  $\Delta \gamma$ for randomly distributed sources also occurs in the [$0-10^\circ$] bin but is only around half the size of the peak in the observed $\Delta \gamma$ distribution. From this we can conclude that there is an overall strong tendency for CDSs to originate near the coronal neutral line. \\ 

\indent 

\section{Discussion}
Past studies have tracked only a handful of individual CDS-associated density structures to probes making \textit{in situ} measurements. The studies have found such blobs to be associated with the passage of complex magnetic field structures, arising near the heliospheric plasma sheet, such as twisted magnetic fields reminiscent of flux ropes \citep{2010JGRA..115.4104R} and refolded magnetic field lines \citep{2011ApJ...734....7R}. The reason given for their  association with density increases observed by heliospheric imagers relates to the idea that they become compressed inside SIR/CIRs during their transit to 1~AU, thereby corresponding a local density enhancement.\\
\indent While these studies suggest that CDSs originate in the vicinity of the coronal neutral line, a number of questions remain open: 
\begin{itemize}
\item are all CDSs associated with the magnetic field polarity inversion line?
\item are all CDSs that are detected in the ecliptic plane eventually swept up by high-speed streams?
\end{itemize}   
To answer these questions and gain further insight into heliospheric variability, the most comprehensive analysis to date of the origin, kinematic evolution and arrival of these CDSs has been carried out in this study.\\
\indent CDSs are most clearly imaged in the heliosphere during solar minimum, when the heliospheric images are not too perturbed by the presence of CMEs (Figure~\ref{fig:Nevents}). Their derived speeds correspond to that of the slow solar wind throughout the solar cycle (Figure~\ref{fig:histogram_SpeedBeta}; see also \cite{2015SoPh..tmp..111C}). This is in good agreement with the results of analysis of their source regions using a PFSS model, if we relate large expansion factors to slow solar wind speeds (Figure~\ref{fig:stackplot}) as is commonly done \citep{1992ApJ...392..310W,2002JGRA..107.1319A}. Comparison with \textit{in situ} data shows that CDS speeds are not those of the stream interface measured at 1~AU (Figure~\ref{fig:compare_speed}), nor of the stream interfaces tracked via IPS \citep{2010SoPh..261..149B}. We note that IPS is sensitive to density fluctuations and therefore particularly sensitive to the shear regions generated at the stream interface. We conclude that, for most of their propagation, CDSs are therefore associated with density variations in the background slow solar wind. Recent analysis of \textit{Helios} observations near 0.3~AU has revealed that  the densest solar wind is associated with the heliospheric plasma sheet and propagates with a speed of less than 300 km~s$^{-1}$ in over 8\% of all observations \citep{Sanchez-Diaz}. Additionally, SIR/CIRs are rarely well formed at these heights. Therefore, we conclude that the CDSs tracked in this study must undergo little compression, and therefore little acceleration, below 0.3 to 0.5~AU (roughly halfway through the field of view of SECCHI imagers). However, Figures \ref{fig:global_ST-A} and \ref{fig:time_Delays} show that, by the time they reach 1~AU, CDSs tend to arrive at any observing spacecraft close to the peak in density associated with SIR/CIR passage. We hence conclude that between 0.5 and 1~AU, the CDSs are rapidly caught up by ensuing high-speed streams. There is some evidence of this in panels (c) and (f) of Figure~\ref{fig:stackplot} with the clearest velocity gradient occurring between Carrington longitudes of 150 and 300$^\circ$ and Carrington rotation numbers of 2055 and 2090. Panel (f) shows that the CDS source regions tend to occur ahead of (\textit{i.e.} at larger Carrington longitudes than)  the large gradient in speed (shown in black), whereas at the time of arrival at 1~AU, the CDSs have been caught up, and become entrained, by the high speed flow and hence lie inside the regions of large speed gradients.  \\
\indent  Our work has demonstrated that CDSs are mostly associated with magnetic polarity inversions both at the source surface neutral line (Figure~\ref{fig:stackplot}b) and \textit{in situ} at the heliospheric current sheet (Figure~\ref{fig:stackplot}a, Figure~\ref{fig:histoneutral}). This confirms the associations inferred in previous studies based only a handful of events. Hence we conclude that, on average, CDSs originate in the slow solar wind near the source surface neutral line. It is therefore likely that they originate near the heliospheric plasma sheet that commonly forms in this region of the corona. As they propagate to 1~AU, the small-scale plasma blobs of which they are formed get swept up, and hence compressed, by high-speed streams, thereby enhancing their visibility. Their impact at 1~AU is associated with the passage of SIR/CIRs because they are caught up by high-speed flow somewhere between 0.5 and 1~AU.\\


\section{Conclusions}
In this article, we have presented a new catalogue of bright (hence dense) structures that are corotating in the field of view of the heliospheric imagers onboard the STEREO-A (ST-A) satellite. HI images provide a unique way to capture the behavior of such density structures from as close as $20~R_\odot$ from the Sun, out to 1~AU, and beyond. The emergence of transients that subsequently become entrained by SIR/CIRs, is clearly evidenced by running-difference heliospheric images; such behavior produces a characteristic converging signature in a ST-A HI J-map; we term these structures as corotating density structures (CDS). A list of 190 CDS events was generated, covering the period of time from the start of the science phase of the STEREO mission (April 2007)  to the start of reduced operations in August 2014. This covers a solar minimum period (2007--2009) and a part of solar maximum (2012--2014) and the ascending phase in between.
Our results show that, under quiet solar conditions, heliospheric imaging can be used to systematically track all CDSs but that, solar condition dependent, our ability to identify CDSs is severely hampered by the perturbation of the heliospheric imagery by frequent CME activity. \\

\indent The average speed of our catalogued CDSs is found to be $311 \pm 31$ km~s$^{-1}$. We have showed that the long-term variation in the CDS speed, derived from HI imagery on ST-A at least, follows closely the speed of the slow solar wind ahead of SIR/CIR compression regions. We have also found that the predicted arrival time of these structures at 1~AU has an accuracy of about 9 hours for ST-A and 11 hours for ST-B. Since the completion of this work, a paper by \cite{2015SoPh..tmp..111C} has been published, which studies a smaller sample of 40 CDSs (although obviously they did not use that terminology) also observed in ecliptic J-maps from HI on ST-A between 2007 and 2010. Those authors come to similar conclusions about the propagation speed of CDSs detected by HI being close to the slow solar wind speed. They also show that accounting for the time-varying separation in ecliptic longitude between the spacecraft and the outflowing transient ($\phi$) in the fitting procedure -- in essence taking into consideration the longitudinal drift of the STEREO spacecraft (Equation~2 of their article) -- improves the speed predictions by a few tens of km~s$^{-1}$ (Figure 6 of their article). In the present work, this effect was not taken into account and, hence, our technique tends to predict speeds some 20-30~km~s$^{-1}$ slower than the \textit{in situ} solar wind values (see Figures~\ref{fig:global_ST-A} and \ref{fig:global_allYears}) and the speeds derived by \cite{2015SoPh..tmp..111C} where the same events were considered. We note that the main conclusions presented in this paper are not significantly affected by the non-implementation of this correction (see comparison with \textit{in situ} catalogue, Section~\ref{subsect:insitu_compar}), but the correction for the angular motion of the satellite will be implemented in upcoming updates to the catalogue. \\

\indent The catalogue was used to test the association between the CDS source regions at the Sun, and their \textit{in situ} arrivals, and the heliospheric current sheet. This analysis demonstrated a strong tendency for the individual density structures (blobs) comprising the CDS to originate near the coronal neutral line and the heliospheric current sheet. In a future study, we will consider that subset of our catalogue where a blob is propagating directly towards an \textit{in situ} observatory spacecraft in order to analyse, in more detail, the in-situ signatures of CDS-associated blobs. In particular, it would be interesting to see whether they are all associated with complex magnetic fields as found in the case studies by \cite{2009ApJ...702..862T} and \cite{2010JGRA..115.4103R}. This will provide additional insight into the variability of the slow solar wind. Finally, we note that both \textit{Solar Orbiter} and \textit{Solar Probe Plus} will track these small transients with much higher temporal and spatial resolution and will likely be able to determine if they form above or below the coronal neutral line located close to the source surface. 
 
%
 \begin{acks}
 This project has received funding from the European Union's Seventh Framework Programme for research, technological development and demonstration under grant agreement No.~606692 (HELCATS).
We thank Yi-Ming Wang for sending us the PFSS computations of the source surface magnetic field and expansion factors used in this paper. We acknowledge usage of the tools made available by the French plasma physics data center (Centre de Donn\'ees de la Physique des Plasmas; CDPP; http://cdpp.eu/), CNES and the space weather service in Toulouse (Solar-Terrestrial Observations and Modelling Service; STORMS). This includes the data mining tools AMDA (http://amda.cdpp.eu/), the CLWEB tool (clweb.cesr.fr/) and the propagation tool (http://propagationtool.cdpp.eu). We acknowledge in particular the support of Christian Jacquey, Myriam Bouchemit and Elena Budnik. VB  acknowledges support from the ``Deutsches Zentrum f\"ur Luft- und Raumfahr (DLR)'' under grant 50OL1201.
 \end{acks}

%
%
 \bibliographystyle{spr-mp-sola}
 \bibliography{biblio}  
\end{article} 
\end{document}